\documentclass[twocolumn]{aastex631}

\usepackage{graphicx}
\usepackage{gensymb}
\usepackage{xcolor} 
\usepackage{amsmath}

\shorttitle{Lunar Micrometeoroid Impact Rate Analysis}
\shortauthors{Yahalomi et al.}
\graphicspath{{./}{figures/}}
\begin{document}

\title{Micrometeoroid Impact Rate Analysis for an Artemis-Era Lunar Base}

\correspondingauthor{Daniel A. Yahalomi}
\email{dyahalomi@flatironinstitute.org}

\author[0000-0003-4755-584X]{Daniel A. Yahalomi} 
\altaffiliation{Flatiron Research Fellow}
\affiliation{Center for Computational Astrophysics, Flatiron Institute, 162 Fifth Ave, New York, NY 10010, USA}
\affiliation{Department of Astronomy, Columbia University, 550 W 120th St., New York NY 10027, USA}

\author[0000-0002-0748-9115]{Matthew T. Scoggins} 
\affiliation{Department of Astronomy, Columbia University, 550 W 120th St., New York NY 10027, USA}

\author{Nasiah Anderson}
\altaffiliation{Columbia STAR Program Student}
\affiliation{Columbia Secondary High School, 425 W 123rd Street, New York NY 10027, USA}

\author{Mark Driker}
\altaffiliation{Columbia STAR Program Student}
\affiliation{Columbia Secondary High School, 425 W 123rd Street, New York NY 10027, USA}

\author{Kokoro Onuma}
\altaffiliation{Columbia STAR Program Student}
\affiliation{Columbia Secondary High School, 425 W 123rd Street, New York NY 10027, USA}

\author{Kwamena T. Awotwi} 
\affiliation{Department of Physics, Columbia University, 550 W 120th St., New York NY 10027, USA}

\author{Justin M. Donovan} 
\affiliation{Department of Mechanical Engineering, Columbia University, 500 West 120th St., New York, NY 10027, USA}

\author{Priyan Sathianathan} 
\affiliation{Department of Industrial Engineering \& Operations Research, Columbia University, 550 W 120th St., New York NY 10027, USA}



\begin{abstract}
NASA's Artemis Mission aims to return astronauts to the Moon and establish a base at the lunar south pole. A key challenge is understanding the threat from micrometeoroid impacts, which are too small to monitor directly. Using NASA's Meteoroid Engineering Model 3 (\texttt{MEM~3}), we estimate micrometeoroid impact rates on a base comparable in size to the International Space Station (100\,m $\times$ 100\,m $\times$ 10\,m). We find that a lunar base would experience $\sim$15,000--23,000 incident impacts per year by micrometeoroids with a mass range of $10^{-6}$--$10^{1}$~g, depending on location -- with minima at the lunar poles, a maximum near the sub-Earth longitude, and a factor of $\sim$1.6 variation between the two. To assess the mitigating effect of protection systems, we present a functional relationship describing the number of impacts that penetrate the shielding as a function of the minimum meteoroid mass capable of penetrating the shield -- the ``critical mass.'' We estimate that state-of-the-art Whipple shields protect against $\sim$99.9997\% of micrometeoroids. By re-running \texttt{MEM~3} with a minimum mass equal to the critical mass of modern Whipple shields, we determine that a shielded lunar base would experience $\sim$0.024--0.037 penetrating impacts per year -- again with minima at the poles and a maximum near the sub-Earth longitude. These results indicate that (1) the lunar poles are optimal for sustained habitation, (2) gravitational focusing by Earth dominates over its geometric shielding for this micrometeoroid flux, and (3) current shielding technology can reduce micrometeoroid threats by nearly five orders of magnitude, making long-duration lunar habitation feasible.


\end{abstract}


\section{Background and Motivation} 

The Artemis program, led by National Aeronautics and Space Administration’s (NASA), marks a renewed commitment to sustained human presence on the Moon. Building upon the legacy of Apollo, Artemis aims not only to return astronauts to the lunar surface but also to establish a long-term base of operations at the lunar south pole. This initiative, in collaboration with international and commercial partners, envisions a new era of lunar exploration that will serve as a stepping stone for missions to Mars and beyond. As planning for surface infrastructure advances, assessing the environmental risks faced by long-duration lunar assets becomes critical. 

Key to this vision is the \textit{Artemis Base Camp} architecture. The base camp concept frames how future landers, habitats, logistics, and operations might evolve on the lunar surface. To design for longevity, one must account for the myriad environmental hazards that a long-duration outpost will face -- among them radiation, extreme thermal cycling, regolith dynamics, seismic shaking, dust, and, of particular importance to this work, impacts.

Artemis~III, currently planned for $\sim$2027, will be an exploratory mission to the lunar south pole, providing reconnaissance for future sustained surface operations. NASA has identified the south polar region -- specifically the \textit{Artemis Exploration Zone} (AEZ) -- as the prime target due to its scientific value and proximity to Permanently Shadowed Regions (PSRs) that may contain accessible water ice \citep{penaasensio_2024}. Building on this framework, \citet{penaasensio_2024} used a multi-criteria decision analysis to identify the Nobile Rim region as a leading candidate for the Artemis~III landing site. The selection criteria encompass features such as stable, flat terrain to ensure safety during landing and operations; unobstructed communication links with Earth to facilitate effective data transfer and mission management; sufficient solar illumination to support power generation; and environmental conditions that keep equipment within acceptable temperature ranges -- all aimed at achieving the highest possible scientific yield \citep{penaasensio_2024}.

While landing site selection focuses on ensuring the safety and scientific value of surface operations, long-term mission success will also depend on protecting habitats, vehicles, and equipment from the harsh lunar environment. One critical consideration in this context is shielding against micrometeoroid and orbital debris (MMOD) impacts, which pose a persistent hazard to both crewed and uncrewed systems.

\subsection{Current MMOD Shielding} 
\label{Current MMOD Shielding}

\begin{figure}[htb!] 
\center
\includegraphics[width=\columnwidth]{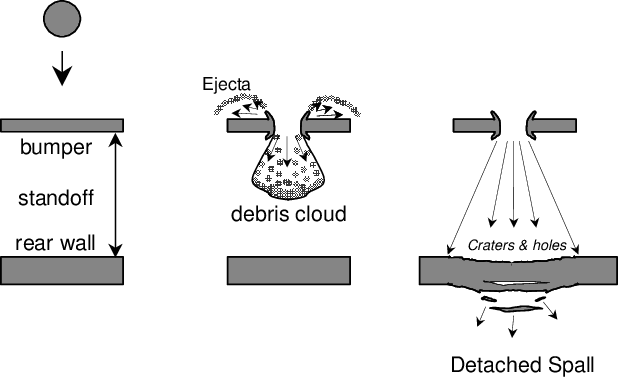}
\caption{Whipple shield concept from \citet{christiansen2003}.}
\label{fig: whipple_shield}
\end{figure}

In establishing a lunar base as a part of the Artemis Mission, there will inevitably be some meteoroid shielding protocol. The specific shielding plans for the Artemis mission are not yet known; however, reasonable inferences can be made based on NASA’s current Micrometeoroid and Orbital Debris (MMOD) shielding designs. The classic MMOD shield (Whipple shield) is composed of a  thin sacrificial bumper and a rear wall, with some interior spacing  \cite{christiansen_nagya_lear_prior_2009} typically constructed of aluminum \citep{christiansen_nagya_lear_prior_2009}. Such a configuration is shown in Figure~\ref{fig: whipple_shield}. Assuming that NASA will use a Whipple shield made of aluminum throughout the Artemis mission, we can estimate the minimum size of a projectile that would be capable of penetrating through the shield. We can use the equation for critical diameter for a Whipple shield, in high velocity space, as presented in \citet{ryan_christiansen_2010}, and shown below:

\begin{equation} \label{eq:ballistic_eq}
    d_c = 3.918 \, F_2^* \, \frac{t_w^{2/3} \, S^{1/3} \, (\sigma/70)^{1/3}}{\rho_p^{1/3} \, \rho_b^{1/9} \, (V \cos \theta)^{2/3}}.
\end{equation}

Here, F$_2^*$ is the projectile fragmentation efficiency, t$_w$ is the thickness in cm of the rear wall, S is the spacing in cm of the rear wall, $\sigma$ is the rear wall yield stress in ksi, $\rho_p$ is the density of the projectile,  $\rho_b$ is the density of the front bumper, V is the velocity of the projectile, and $\theta$ is the angle of impact.

Assuming the projectile is a solid sphere of density, \( \rho_p \), the mass is:

\begin{equation}
    m_c = \frac{\pi}{6} \rho_p d_c^3.
\end{equation}

Substituting Equation~\eqref{eq:ballistic_eq} into the expression for mass, we can determine the critical mass of micrometeoroid impactor:

\begin{equation} \label{eq: critical mass}
    m_c = \frac{\pi}{6} \, (3.918 \, F_2^*)^3 \, \frac{t_w^2 \, S \, (\sigma/70)}{\rho_b^{1/3} \, (V \cos \theta)^2} .
\end{equation}

Using this equation and representative parameter values (see, e.g., \citealt{ryan_christiansen_2010}), a back-of-the-envelope calculation for the fastest and most dense micrometeoroids -- with impact velocities up to 72~km/s and densities up to $8.90$~g~cm$^{-3}$ -- yields a critical shield diameter of approximately 0.12~cm. This value should be regarded as an order-of-magnitude estimate, indicating that objects equal to or larger than roughly 0.12~cm in diameter may exceed the protective capability of state-of-the-art MMOD shields. 

This underscores the importance of precise and accurate modeling of the micrometeoroid environment. It is worth stating that micrometeoroids with these extreme impact speeds ($>$50 km/s) are likely retrograde cometary micrometeoroids that have been circularized by P-R drag near 1 au, and thus likely have lower densities than is assumed here (see, e.g., \citealt{Nesvorny2011}). However, we chose to be maximally conservative in our shielding estimates throughout to ensure our shielding analysis captures the upper bound of potential impact severity.

We also note that NASA may ultimately employ novel shielding strategies -- such as the use of lunar regolith in shielding, which will be abundant on the surface -- but, no specific plans or experimental data on their performance have yet been disclosed. Accordingly, in what follows we proceed under the assumption of a Whipple-type shielding configuration as a representative baseline.

\subsection{Current Lunar Impact Monitoring}
Current lunar impact monitoring techniques employ several complementary observational strategies, each with distinct strengths and limitations. These include topographic mapping through laser altimetry, detection of impact-induced optical flashes on the lunar surface, temporal imaging of newly formed craters, and scaling from bolide detection on Earth. Collectively, these methods have greatly advanced our understanding of the lunar impact environment, particularly within the meteoroid size regime where optical and morphological signatures are more readily detectable. However, even within this observable size range, significant uncertainties remain in quantifying the present-day meteoroid impact flux on the lunar surface. As noted by \citet{Speyerer2016}, ``although studies of existing craters and returned samples offer insight into the process of crater formation and the past cratering rate, questions still remain about the present rate of crater production.'' 

In the following, we summarize the principal observational approaches, and findings, currently used to monitor meteoroid impacts on the Moon.

\begin{itemize}
    \item \textbf{Temporal Imaging:} The Lunar Reconnaissance Orbiter Camera (LROC), launched in 2009, captures high-resolution images of the Moon's surface, enabling the detection of new impact craters through temporal image comparisons. \citet{speyerer2016quantifying} identified over 200 new craters, with a resolution limit of approximately 10 meters. Laboratory experiments and dimensional analyses show that crater size depends on impactor size, velocity, gravity, and material properties in a well-characterized scaling framework \citep[see e.g.,][]{melosh1989impact,holsapple1993scaling}. For meteor impactors striking the lunar surface at typical impact velocities, the final crater diameter is typically $\sim$10--20 times larger than the projectile diameter. Thus, the LROC temporal imaging data is sensitive to craters corresponding to meteors roughly 0.5 to 1 meters in diameter. Therefore, this technique is insensitive to smaller, micrometeoroid impacts and is additionally limited by the availability and cadence of suitable image pairs

    \item \textbf{Flash Detection:} Earth-based telescopes monitor the Moon’s nearside hemisphere for brief optical flashes produced by hypervelocity meteoroids striking the lunar surface. Observations are constrained to nighttime, favorable weather, and specific lunar phases when illumination is less than 50\%, which optimize contrast against the dark background. The observed flash brightness, often modeled as black-body emission, is assumed to represent a fraction (the luminous efficiency) of the impactor’s kinetic energy; with an estimate of impact velocity, this enables derivation of the meteoroid’s mass and size. Over the past decades, this technique has allowed determination of the flux and size distribution of small near-Earth objects in the centimeter regime \citep{madiedo2015perseid, Avdellidou2019}. The Lunar Meteoroid Impact Observer (LUMIO) is a CubeSat scheduled to launch in 2027 in order to observe the lunar farside for light flashes produced by impacts. By operating at the Earth–Moon L2 point, LUMIO's observations are not limited by weather and it eliminates noise from Earth-shine \citep{tos_topputo_2018}. LUMIO's primary science goal is to answer \textit{``what are the spatial and temporal characteristics of meteoroids impacting the Lunar surface?''} and its sensitivity will extend into the micrometeoroid regime \citep{Cervone2022}.

    \item \textbf{Topographic Mapping:} The Lunar Orbiter Laser Altimeter (LOLA) measures elevation changes on the lunar surface via laser altimetry to detect large-scale impact events. LOLA performs optimally in characterizing the topography at the poles where the LRO orbits converge. While highly accurate for broader terrain mapping, LOLA's spatial resolution limits detection to craters larger than approximately 300--400 meters. Via typical impact scaling relations, these craters suggest minimum impactor sensitivity on the order of 15--40 meters.  This makes it currently unsuitable for tracking the small-scale micrometeoroid environment for Artemis-era lunar surface operations \citep{Smith2010, Kereszturi2022}.

    \item \textbf{Bolide observations from Earth:} Another relevant data point for meteor impacts on the lunar surface comes from scaling bolide observations on Earth. Bolides are bright meteors that burn up in the Earth's thick atmosphere. It is estimated that about 100 meteoroids of $\sim$1 m diameter burn-up in Earth's atmosphere per year (see, e.g., \citealt{bolides1,bolides2}). Scaling by the relative sizes, one would expect the Moon to receive roughly one-twentieth as many impacts -- or about five such $\sim$1 m objects per year.

\end{itemize}

These constraints collectively highlight the need for more sensitive modeling of the lunar surface prior to the establishment of a long-term lunar base and long-term human presence on the Moon. Perhaps most strikingly, despite their demonstrated success in characterizing larger impact events, these techniques lack the sensitivity required to systematically monitor or constrain impacts in the micrometeoroid regime.

\subsection{NASA's Meteoroid Engineering Model 3 (\texttt{MEM~3})}
\label{subsec:mem3}

NASA's Meteoroid Engineering Model version~3 (\texttt{MEM~3}) is the agency's current physics-based model of the inner–solar-system meteoroid environment \citep{McNamara_2004, Moorhead2020_MEM3_NTRS}. Given a user-specified, time-dependent trajectory, \texttt{MEM~3} returns velocity-resolved, directional fluxes and a bulk-density distribution for meteoroids in the mass range $10^{-6}-10^{1}$~g encountered along that path, explicitly accounting for \emph{gravitational focusing} and \emph{planetary shielding} near major bodies -- Earth, Moon, Mercury, Venus, and Mars \citep{Moorhead2020_MEM3_NTRS}. These features are essential when translating interplanetary fluxes to the near-Moon environment, where local gravity from the Earth perturbs meteoroid trajectories and the solid body of the Earth occludes part of the sky.

\texttt{MEM~3} adopts the meteoroid mass distribution, as presented in \citet{Grun1985}, using the full analytical form of their flux model (see Equation~\ref{eq:grun}). The flux is scaled to an arbitrary limiting mass using this relation, but its overall amplitude is normalized to observations of the meteoroid influx at the top of Earth's atmosphere from the Canadian Meteor Orbit Radar (CMOR)~\cite{CampbellBrown2008}. In the current implementation, all meteoroid source populations are assumed to follow the same mass distribution.

Compared to prior \texttt{MEM} releases, \texttt{MEM~3} improves the correlation between impact direction and speed, incorporates a bulk-density distribution, updates sporadic source populations, and provides both GUI and command-line interfaces for efficient mission analyses \citep{Moorhead2020_MEM3_NTRS, Moorhead2020_paper}. Model behavior and predicted impact rates have been compared against spacecraft records (e.g., LDEF; Pegasus II/III), demonstrating good agreement -- within a factor of 2-3 of the \texttt{MEM~3} prediction \citep{Moorhead2020_paper}. 

Although \texttt{MEM~3} is often used for Earth-orbiting satellites and interplanetary cruise phases, its directional, velocity-dependent outputs are applicable to a fixed lunar installation by treating the habitat as a ``spacecraft'' with a stationary state vector on the Moon's surface. In that configuration, \texttt{MEM~3} provides the incident flux as a function of local time and look direction at the site of interest, thereby enabling \textit{site-specific} penetration and damage-risk assessments. Beyond engineering usage, \texttt{MEM}-based flux predictions have been leveraged in planetary science contexts (e.g., interpreting Bennu's particle-ejection events), demonstrating the model's relevance to the near-Earth micrometeoroid population \citep{Bottke_2020}.

\section{Methods}
\begin{figure*}[htb!] 
\center
\includegraphics[width=\textwidth]{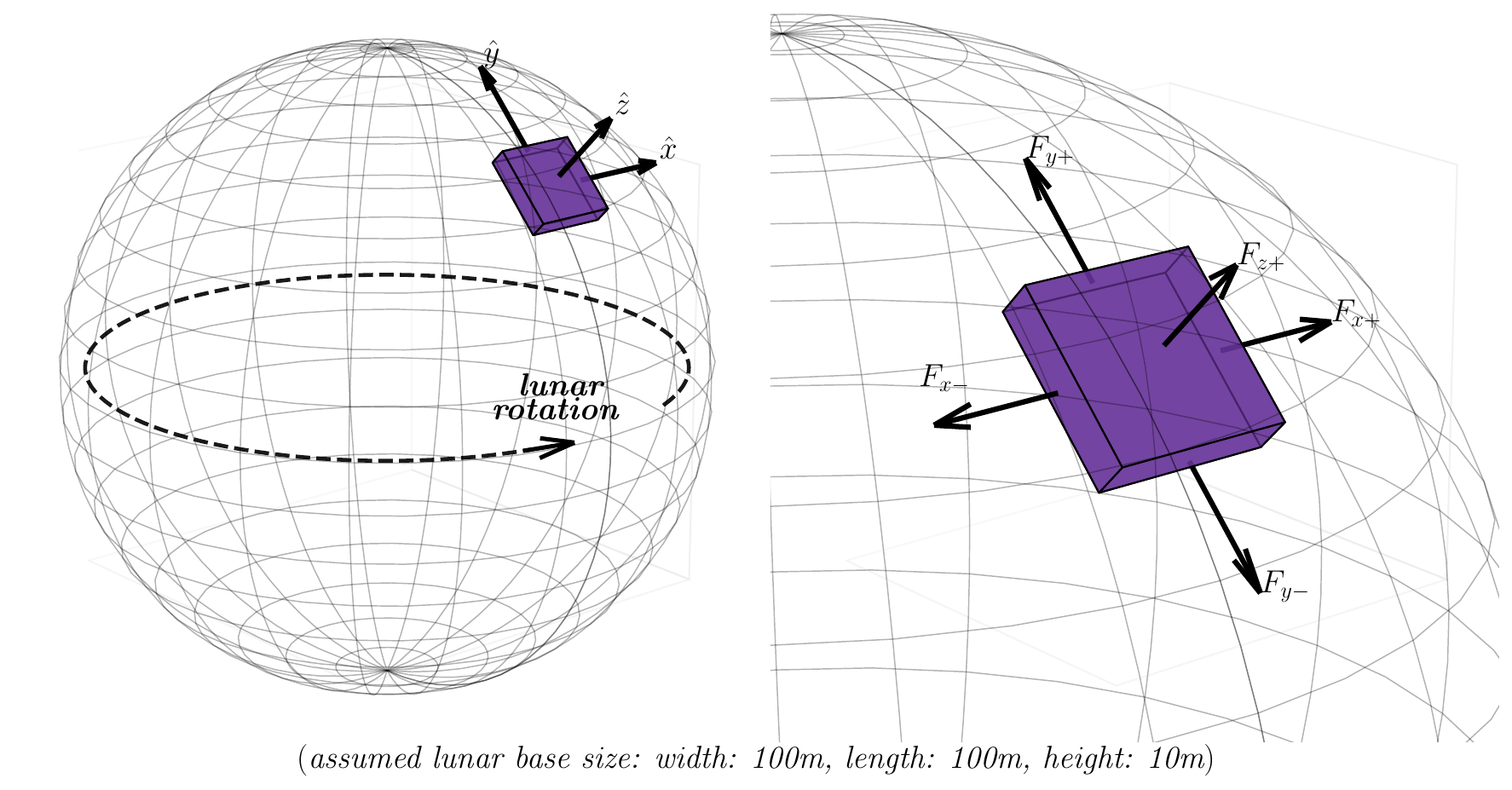}
\caption{Coordinate system used in the \texttt{MEM~3} simulations for [left] full lunar surface and [right] zoom-in on the lunar base.}
\label{fig: schematic}
\end{figure*}

Each run of the \texttt{MEM~3} code provides outputs for one lunar base. To derive the impact rate across the full surface of the Moon, we draw 1,000 points from a Fibonacci sphere to create the initial locations of our lunar base, in selenographic coordinates, described by $\phi$ measured from the lunar north pole and $\theta$ measured from the Moon's prime meridian. Sampling the Fibonacci sphere distributes the bases evenly across the Moon's surface, with a $\phi$ distribution that is approximately normal and a uniform $\theta$ distribution.

All trajectories start at J2000, or the J${\sim}2451544$ Julian date, with $\theta=0$ pointing towards the vernal equinox. We calculate the trajectory of the lunar base for the draconic period of the Moon $T=27.2122$ days, and calculate 30 snapshots of the lunar base over this period for the trajectory file.  \texttt{MEM~3} allows the coordinate system to be centered on the Moon, in either ecliptic or equatorial coordinates. We choose to run the simulations in an ecliptic coordinate frame, which requires a transformation of our selenographic coordinates. Given the initial base location in selenographic coordinates, L($t_0$)$=$[$\phi_0$, $\theta_0$], defined by $\phi_0$ and $\theta_0$ at time J2000, $t_0$, the location at a later time $t$ can be estimated as L($t$)$=$[$\phi_0$, $\theta_0+\omega (t-t_0)$] for angular velocity $\omega = 2\pi/T$. This location is then converted into Cartesian coordinates, L($t$)$=$[x, y, z]. The velocity at a given surface location is calculated as the rotational velocity of the base due to the Moon's rotation. The velocity vector is given by
\begin{equation}
    \mathbf{v}(t) = \frac{V}{|r_{\mathrm{xy}}|}\,[-y,\, x,\, 0],
\end{equation}
where
\begin{equation}
    V = \omega\, r_{\mathrm{Moon}} \sin(\phi_0),
\end{equation}
and
\begin{equation}
    |r_{\mathrm{xy}}| = \sqrt{x^{2} + y^{2}}.
\end{equation}

for Moon radius $r_{\rm Moon}$. With the base location in Cartesian coordinates and the base's velocity, we then transform these into the ecliptic by rotating the coordinate system by the Moon's obliquity relative to the ecliptic plane, a rotation of ${\sim} 1.54\degree$. This is an approximation that ignores the libration of the Moon, which can be up to $7 \degree$. This means our base locations can be off by as much as $200$ km, which is on the order of the separation between our simulated base locations.

\texttt{MEM~3} outputs the flux files with a directional dependent flux. We choose to center our output origin to the Moon, and use the body-fixed axes. In the body-fixed system, the $\hat{x}_+$ direction is always the direction of motion of the spacecraft (in our case, the direction that the Moon is rotating, counterclockwise in the ecliptic plane), and $\hat{y}_+$ is determined by the cross product of $\hat{r}$, the radial vector relative to the Moon's center, with $\hat{x}_+$. $\hat{z}_+$ is then the cross product of $\hat{x}_+$ and $\hat{y}_+$, meaning $\hat{z}_+$ will always point in the radial direction, directly away from the surface of the Moon. This can be seen in Figure~\ref{fig: schematic}.

\texttt{MEM~3} simulates two distinct meteoroid populations: a higher-density component and a lower-density component, denoted here by $F_{\alpha}$ and $F_{\beta}$, respectively. Within each population, the density is assumed independent of speed, direction, and mass. For a given lunar base, the total flux is computed as the sum of contributions from both populations.  

The \texttt{MEM~3} model outputs fluxes in units of $\mathrm{m^{-2}\,yr^{-1}}$. To facilitate comparison with the total flux incident on a specific structure, we convert these to units of impacts per lunar base per year by scaling with the surface area of the base along each plane of impact. Assuming a lunar base roughly the size of the International Space Station, we take its height, length, and width to be
\begin{equation*}
    h = 10~\mathrm{m}, \quad l = 100~\mathrm{m}, \quad w = 100~\mathrm{m}.
\end{equation*}

The total flux from the higher-density population at a specific selenographic initial location defined by $\phi_0$, $\theta_0$ at J2000 is then
\begin{align}
    F_{\alpha}(\phi_0,\theta_0) &= lw\,[F_{\alpha,z_{+}}(\phi_0,\theta_0)] \nonumber \\
    &\quad + lh\,[F_{\alpha,x_{+}}(\phi_0,\theta_0) + F_{\alpha,x_{-}}(\phi_0,\theta_0)] \nonumber \\
    &\quad + lh\,[F_{\alpha,y_{+}}(\phi_0,\theta_0) + F_{\alpha,y_{-}}(\phi_0,\theta_0)],
\end{align}
where $F_{\alpha,i_{\pm}}(\phi_0,\theta_0)$ denotes the higher-density flux incident from the $\pm i$ direction, with $i \in \{x, y, z\}$, expressed in $\mathrm{m^{-2}\,yr^{-1}}$ and integrated over all velocities. 
Each flux term also implicitly depends on the minimum simulated mass threshold $m_{\min}$, denoted as $F(>m_{\min}\,|\,\phi_0,\theta_0)$, but this dependence is omitted here for clarity and conciseness.

Analogously, the total flux from the lower-density population is
\begin{align}
    F_{\beta}(\phi_0,\theta_0) &= lw\,[F_{\beta,z_{+}}(\phi_0,\theta_0)] \nonumber \\
    &\quad + lh\,[F_{\beta,x_{+}}(\phi_0,\theta_0) + F_{\beta,x_{-}}(\phi_0,\theta_0)] \nonumber \\
    &\quad + lh\,[F_{\beta,y_{+}}(\phi_0,\theta_0) + F_{\beta,y_{-}}(\phi_0,\theta_0)].
\end{align}

\begin{figure*}[htb!]
    \centering
    \includegraphics[width=0.49\textwidth]{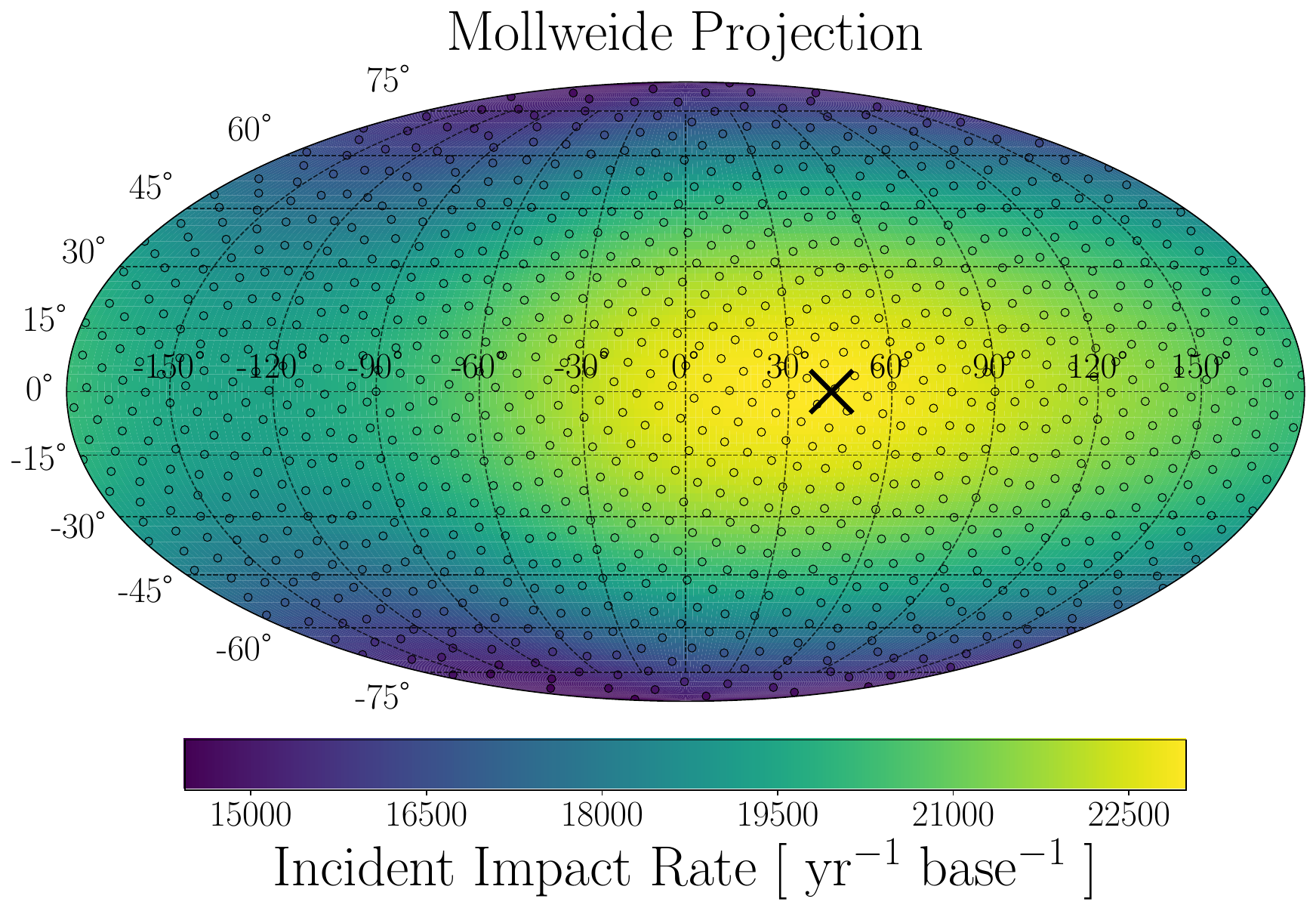}
    \hfill
    \includegraphics[width=0.49\textwidth]{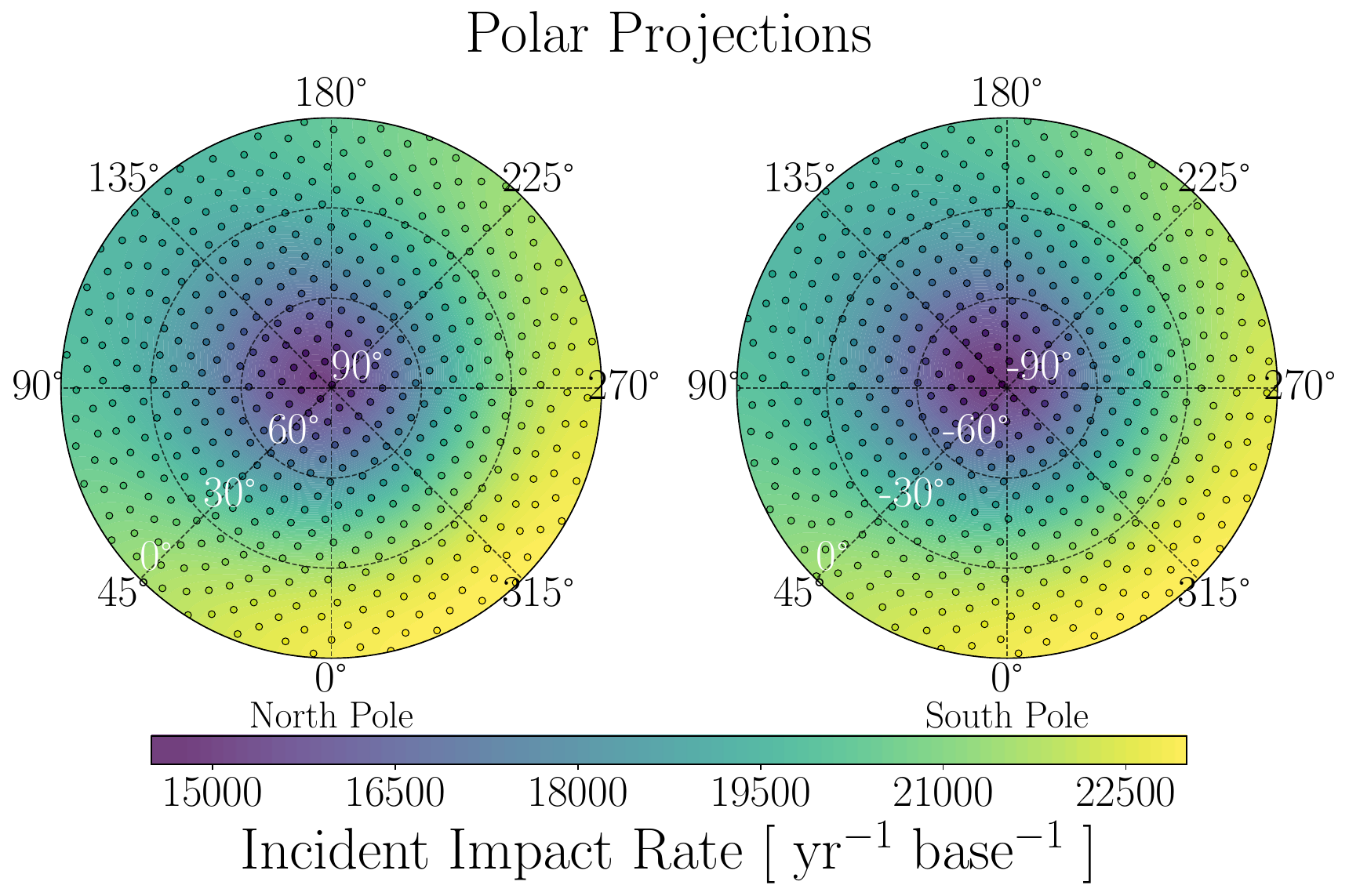}
    \caption{Mollweide (left) and polar (right) projections of the incident micrometeoroid impact rate on the lunar surface, representing the complete $10^{-6}$--$10^{1}$~g mass range in \texttt{MEM~3}. The ``x'' marks the sub-Earth point on the lunar surface.}
    \label{fig:impact_rate_combined}
\end{figure*}

Finally, the total meteoroid flux incident on the lunar base, in units of impacts per year per base, is
\begin{equation}
    F(\phi_0,\theta_0) = F_{\alpha}(\phi_0,\theta_0) + F_{\beta}(\phi_0,\theta_0),
\end{equation}
where all fluxes are implicitly understood as being integrated over masses greater than $m_{\min}$.

\section{Results \& Discussion}

\subsection{Incident Impact Rate}

The \texttt{MEM~3} model allows the user to specify the minimum meteoroid particle mass within the range $10^{-6}$–$10^{1}$~g. As an initial case, we define the incident impact rate as the total micrometeoroid flux on the exterior surface of the base, integrating over this full \texttt{MEM~3} mass range. This represents the raw meteoroid environment -- the rate at which particles would strike the habitat if there was no shielding. Although this configuration is physically unrealistic, since any structural walls of a habitat would provide some degree of protection, it provides a useful baseline from which to quantify the reduction in impact rate achieved by shielding in subsequent analyses.

As previously described, we simulate 1,000 points evenly distributed across the lunar surface as a Fibonacci sphere. This gives a set of discrete impact rates on the lunar surface. To obtain a continuous representation of meteoroid impact rates across the lunar surface, we interpolate the results of these 1,000 discrete \texttt{MEM~3} simulations sampled at different selenographic latitudes and longitudes. The interpolation is performed using a radial basis function (RBF) scheme implemented in the \texttt{SciPy} library \citep{Virtanen2020}. In this approach, each simulated point $(\phi_i, \theta_i)$, corresponding to latitude and longitude, is associated with a total impact rate $F_i$ derived from \texttt{MEM~3}. The function constructs a two-dimensional interpolant
\begin{equation}
    F_{\mathrm{interp}}(\phi, \theta)
    = \sum_{i=1}^{N} w_i\,\varphi(r_i),
\end{equation}
where $r_i = \sqrt{(\phi - \phi_i)^2 + (\theta - \theta_i)^2}$ is the great-circle distance (in degrees) between evaluation and sample points, $\varphi(r_i)$ is the chosen radial basis function, and $w_i$ are the weights obtained by solving the linear system enforced by the known values $F_i$.  

We employ the ``multiquadric'' kernel,
\begin{equation}
    \varphi(r) = \sqrt{1 + (\epsilon r)^2},
\end{equation}
which provides smooth global interpolation suitable for data on a spherical surface. A small smoothing factor (\texttt{smooth=1}) is applied to mitigate overfitting due to local fluctuations in the discrete model output.

This results in the impact rate maps shown in Figure~\ref{fig:impact_rate_combined} -- for Mollweide and polar projections, respectively. The sub-Earth point (“x”) on the lunar surface was computed at the J2000 epoch using \texttt{Astropy} \citep{astropy:2013, astropy:2018, astropy:2022} with the DE432s ephemeris. The barycentric positions of the Earth and Moon were used to form a Moon–Earth vector in the ICRS frame, from which the sub-Earth longitude was obtained via $\theta = \tan^{-1}(y/x)$ and wrapped to $[-180^\circ, +180^\circ]$ for Mollweide map projection compatibility. 

Three key trends emerge from this incident impact rate map: (1) impact rates are high, ranging from approximately 15,000 to 23,000 impacts per year depending on base location; (2) the lunar poles experience systematically fewer impacts than the equatorial regions; and (3) Earth’s gravitational focusing dominates over its shielding effect, with the maximum impact rate occurring near the sub-Earth longitude (see \citet{Moorhead2020_grav_focusing} for further discussion of planetary gravitational focusing of meteor streams).


\subsection{Penetrating Impact Rate}

\begin{figure*}[htb!]
\gridline{
    \fig{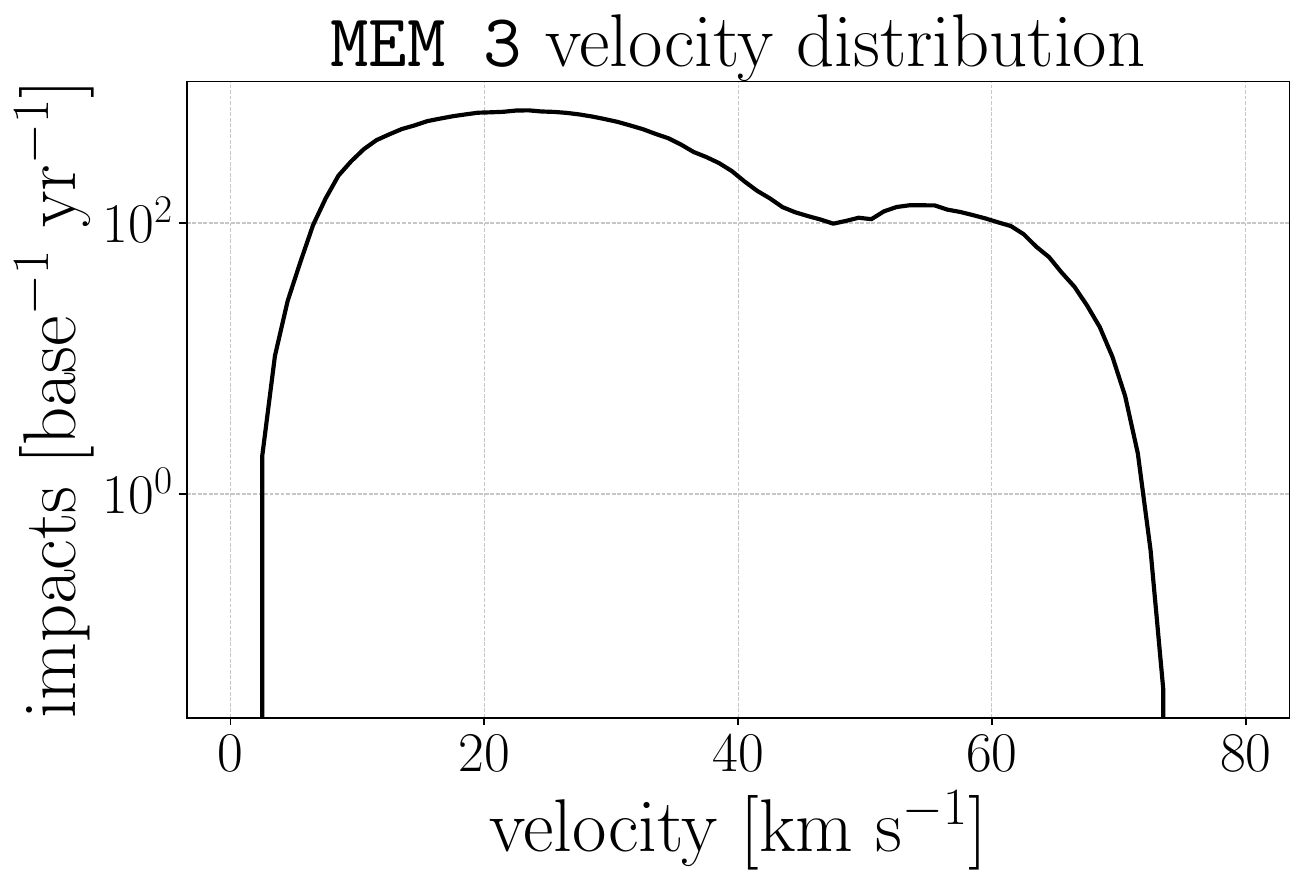}{0.32\textwidth}{(a) Averaged velocity (over all base locations) distribution for \texttt{MEM~3} with $m_{\rm min} = 10^{-6}$g.} 
    \fig{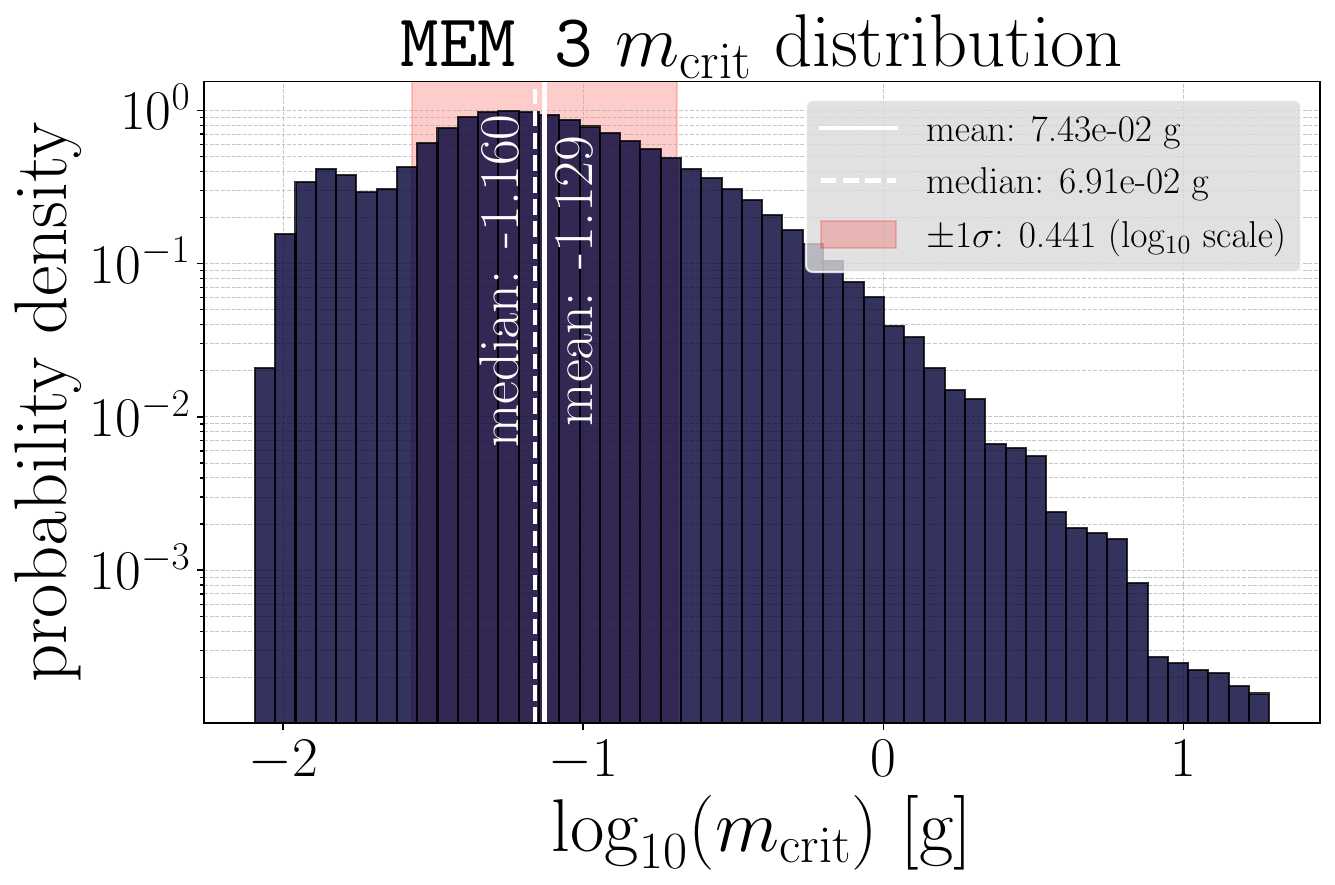}{0.32\textwidth}{(b) The $m_{\rm crit}$ distribution using our velocity distribution and Equation~\ref{eq: critical mass} for the most dense micrometeoroids.}
    \fig{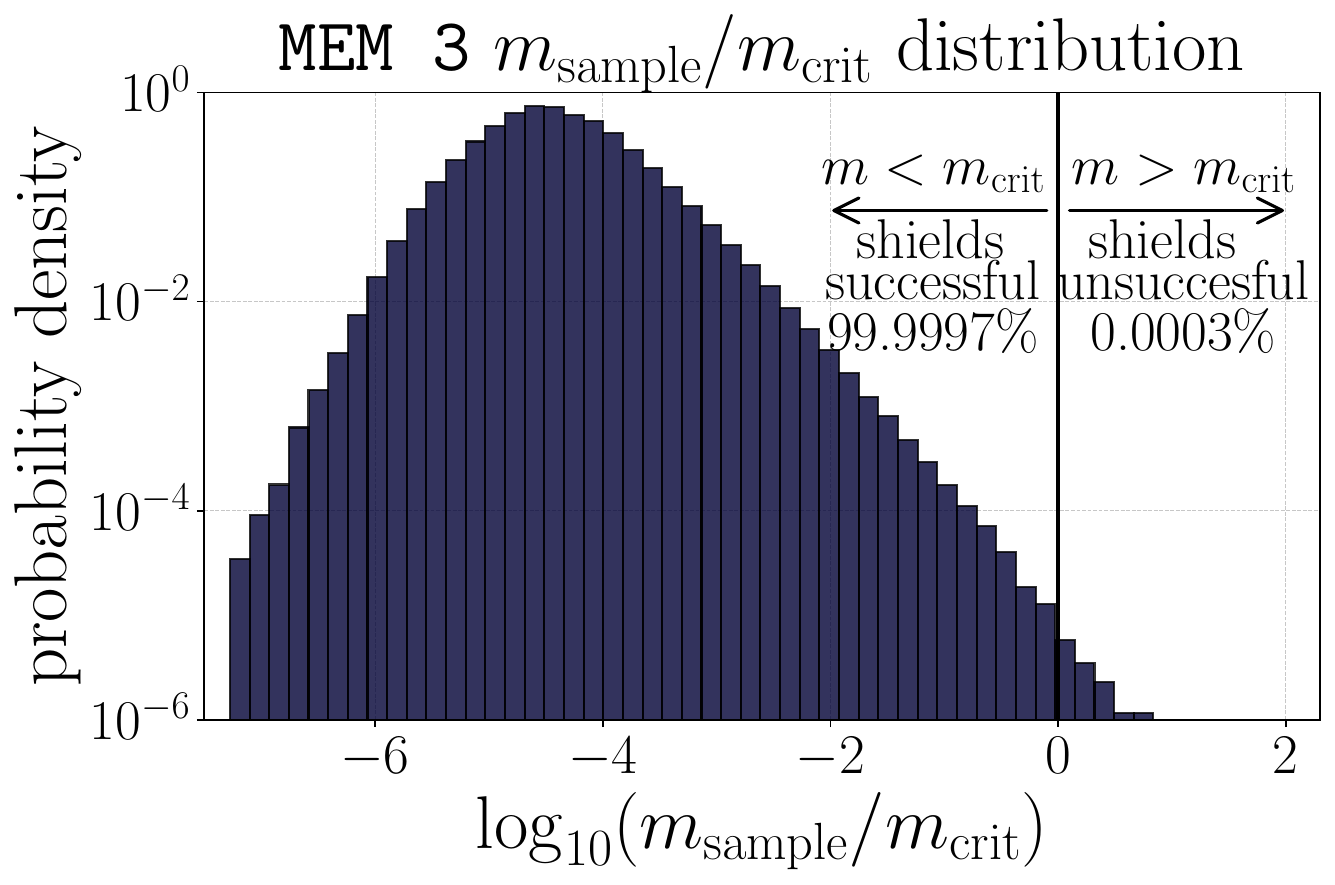}{0.32\textwidth}{(c) The ratio between random samples from mass distribution and $m_{\rm crit}$.}
}
\caption{Impact velocity averaged over all base locations; the derived $m_{\rm crit}$; and the $m_{\rm sample}$/$m_{\rm crit}$ distributions.}
\label{fig:distributions}
\end{figure*}

As discussed previously, the performance of a meteoroid shield can be characterized by its critical mass, $m_c$ (Equation~\ref{eq: critical mass}), defined as the minimum projectile mass that the shield cannot prevent from penetrating. To evaluate this parameter, we adopt a conservative approach to the ballistic limit function, selecting input values that minimize the critical mass and thereby represent a worst-case scenario. This ensures that if a shield design is predicted to withstand an impact, the assessment remains robust under the most adverse, but possible, conditions.

For simplicity, we assume $F_2^* = 1$, which is appropriate for projectiles below the critical mass of a Whipple shield \citep[see][]{ryan_christiansen_2010}. We further assume a normal incidence angle ($\theta = 0$) to yield the smallest critical mass. Additional parameters are adopted from \citet{ryan_christiansen_2010}: rear wall thickness $t_w = 0.48$~cm, bumper density $\rho_b = 2.851$~g~cm$^{-3}$, rear wall yield stress $\sigma = 52$~ksi, and rear wall spacing $S = 11.43$~cm. 

Meteoroids are composed primarily of silicate minerals (Si and O), though metallic constituents such as Fe and Ni are also common \citep{Jessberger1988,Love1993,Flynn2016}. To maintain a conservative estimate of impact severity, we assume a nickel composition for the projectile, corresponding to a density of $\rho_p = 8.90$~g~cm$^{-3}$. We again highlight that this is a very conservative bulk density assumption, as most impactors will likely be comet dust, based on studies of lunar regolith. Decreasing the assumed bulk density would result in \textit{less} predicted impacts penetrating shielding.

We then are left with the velocity of the impactor as the only remaining parameter in our critical mass equation. For the simulation suite with the lowest minimum particle mass ($m_{\rm min} = 10^{-6}$ g), we compute the velocity distribution using the \texttt{MEM~3} model, weighted by directionality according to the assumed base geometry, and then average the resulting flux over each lunar base location. Our averaged velocity distribution is calculated as the flux from both the high and low density contributions, and is shown in Figure~\ref{fig:distributions}. 



From this velocity distribution, we compute the normalized cumulative distribution, which allows us to randomly draw $10^7$ samples. Plugging in this distribution into Equation~\ref{eq: critical mass} we get a distribution of critical masses based on current Whipple shield capabilities, as shown in Figure~\ref{fig:distributions}. The median critical mass of this distribution is $m_{\rm crit} = 10^{-1.16}$g. 

We define the penetrating impact rate as the total micrometeoroid flux that would penetrate through lunar base shielding. We can evaluate the penetrating impact rate in two ways: (1) compare a random sample of impacts from our \texttt{MEM~3} simulation to the critical mass to determine what fraction of impacts are larger than our median critical mass and (2) re-run the \texttt{MEM~3} simulation with the minimum mass equal to the median critical mass, $m_{\rm min} = m_{\rm crit} = 10^{-1.16}$ g.

\begin{figure*}[htb!]
    \centering
    \includegraphics[width=0.49\textwidth]{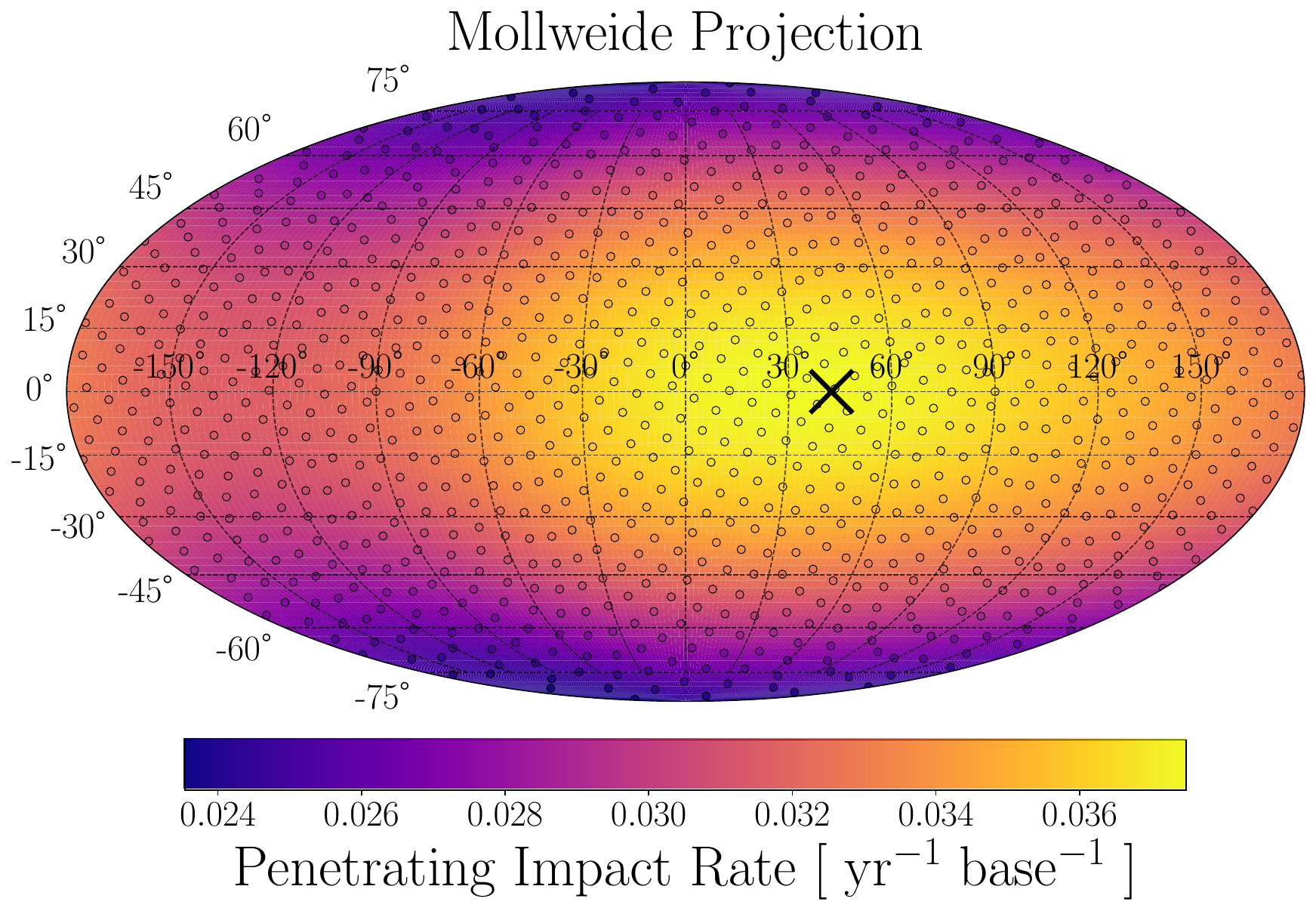}
    \hfill
    \includegraphics[width=0.49\textwidth]{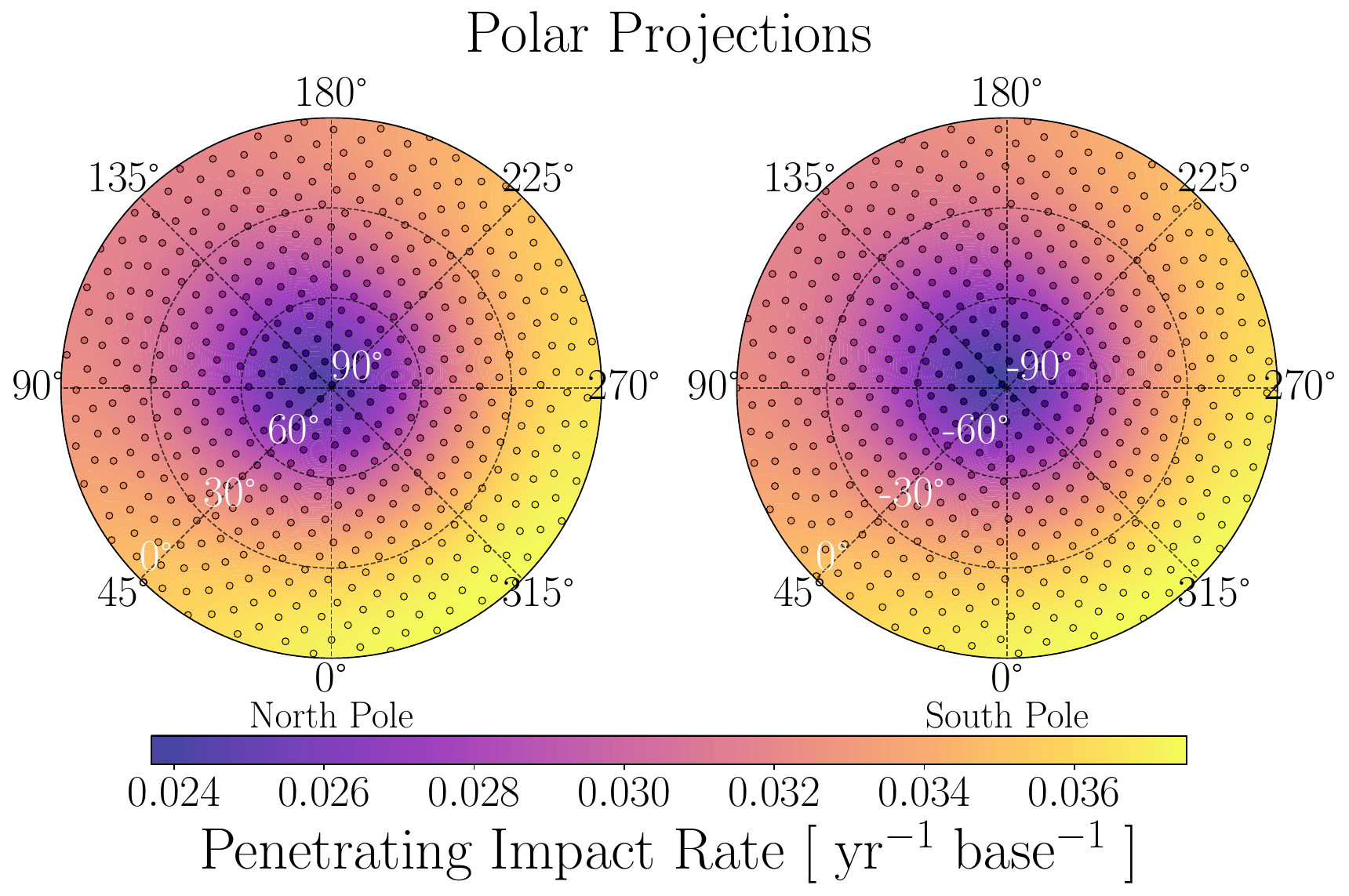}
    \caption{Mollweide (left) and polar (right) projections of the penetrating micrometeoroid impact rate on the lunar surface, representing the $10^{-1.16}$--$10^{1}$~g mass range in \texttt{MEM~3}, for which the most dense micrometeoroids would penetrate current Whipple shielding. The ``x'' marks the sub-Earth point on the lunar surface.}
    \label{fig:shielded_combined}
\end{figure*}

The ratio of randomly sampled masses from our mass distribution compared to the derived critical is shown in Figure~\ref{fig:distributions}. We find that $99.9997\%$ of these particles have masses below the critical mass. Given that this is calculated with the average velocity distribution, a base on the Moon located in an area experiencing fewer than the average number of impacts (i.e. near the poles) and accounting for the fact that most impacts will not be face-on ($\theta >0$), this implies that current shielding is capable of protecting against nearly every micrometeoroid impact. From our simulations, we would expect $\sim$15,000 total annual meteoroid impacts on the lunar base -- however these results suggest that $99.9997\%$ of these micrometeoroids will not get through the shielding. Thus, there will be $\sim$0.045 impacts per base per year that can potentially break through the shielding. This implies that only once every $\sim22$ years will an impact break through Whipple shielding.

We also re-run our full \texttt{MEM~3} model, but with a minimum mass set to the median critical mass, $m_{\rm min} = m_{\rm crit} = 10^{-1.16}$ g. This results in the \textit{penetrating impact rate} map, as shown in Figure~\ref{fig:shielded_combined} -- for Mollweide and polar projections, respectively. 

Interpreting the penetrating impact rate map, we again find that the lunar poles are impacted systematically less than the equator and that the gravitational focusing from Earth dominates over its planetary shielding as the maximum impact rate occurs at the location of the Earth in the lunar sky. With the updated minimum mass set equal to our estimated critical mass, we find that a lunar base will be impacted $\sim$0.024--0.037 per year based on its location. At the poles, we estimate that there will be $\sim$0.024 impacts per year large enough to break through Whipple shielding or once every $\sim42$ years.

We quickly note that since our impact simulations use nominal \texttt{MEM~3} fluxes, they do not consider events such as the destruction of a satellite in the near-Moon environment (or other similar events) that could dramatically increase the impact flux rate over a short timescale.

\subsection{Varying Shielding Capabilities}

Using the \texttt{MEM~3} simulation suite, we can estimate the number of meteoroid impacts at the lunar south pole as a function of the critical shielding diameter. This provides a framework for determining the number of \textit{penetrating} impacts—those that penetrate the protective layers—as a function of the shield’s critical performance threshold. As Artemis-era surface habitat designs mature, such calculations can help identify the critical impact mass or size that shielding must withstand to meet mission safety requirements.

\begin{figure*}[htb!] 
\center
\includegraphics[width=\textwidth]{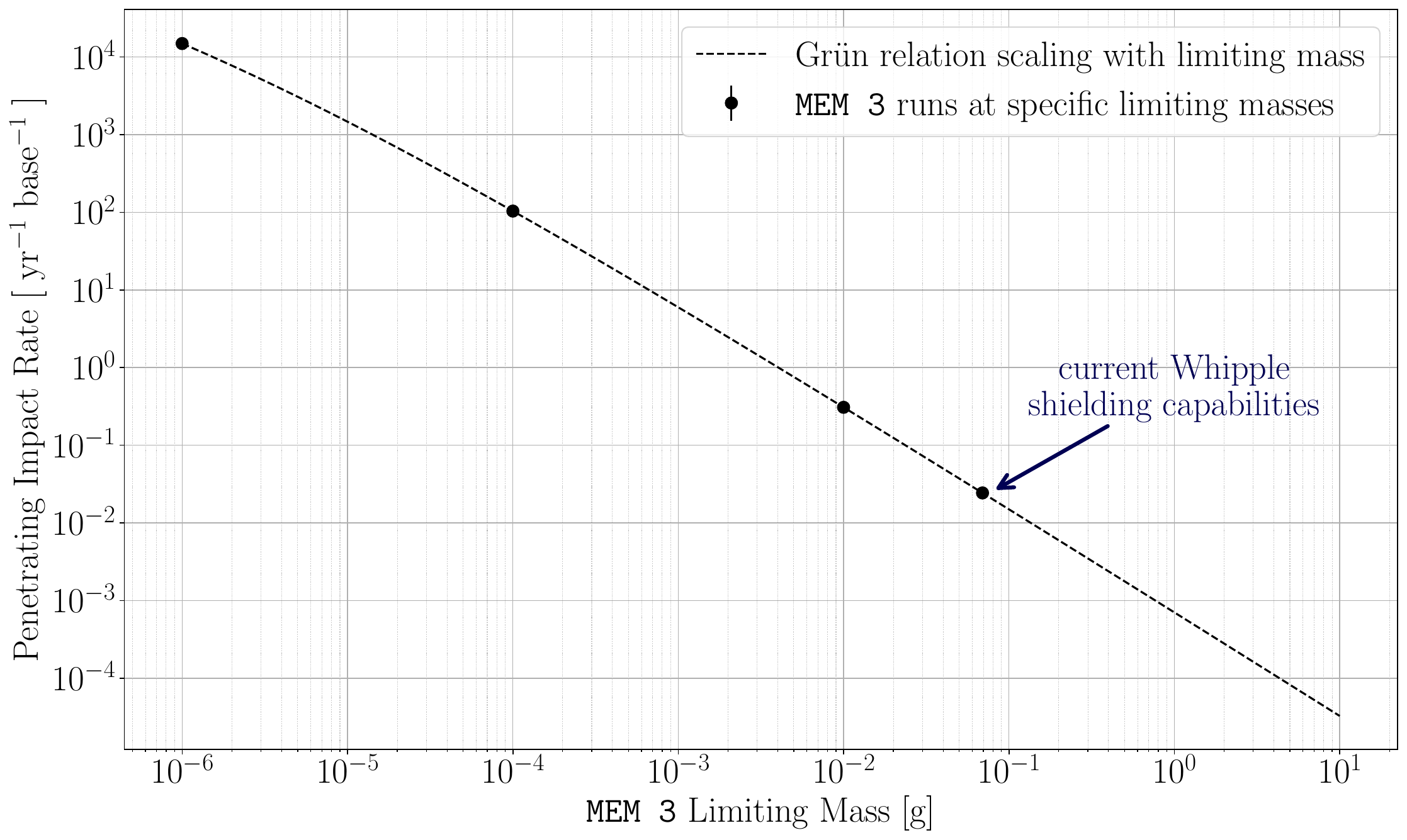}
\caption{Impact rate on the lunar south pole (mean of simulations within $6^{\circ}$ on lunar south pole) as a function of minimum \texttt{MEM~3} mass. Points represent 4 individual \texttt{MEM~3} runs with minimum mass set to $10^{-6}$, $10^{-4}$, $10^{-2}$, $10^{-1.16}$ g, respectively. Dashed line shows the Gr\"un relation scaled to the $10^{-6}$ g minimum mass \texttt{MEM~3} simulation.}
\label{fig: grun}
\end{figure*}

To quantify this, we run two additional \texttt{MEM~3} suites with minimum particle masses set to $m_{\rm min} = 10^{-2}$~g and $m_{\rm min} = 10^{-4}$~g, complementing our previous simulations at $10^{-6}$~g and $10^{-1.16}$~g. In total, these four \texttt{MEM~3} simulation sets provide impact fluxes $F(>m_{\rm min}|\phi_0,\theta_0)$ for different minimum masses $m_{\rm min}$ at 1,000 locations of $(\phi_0,\theta_0)$ across the lunar surface. For each set, we calculate the mean impact rate across all locations within $6^{\circ}$ of the lunar south pole, representing the expected flux at a notional base site. The resulting mean southern-pole impact rates $F(>m_{\rm min}|\phi_{\rm south},\theta_{\rm south})$ are plotted in Figure~\ref{fig: grun} as a function of their respective \texttt{MEM~3} limiting mass $m_{\rm min}$.

To provide a smooth, physically motivated comparison, we also compute the expected mass-dependent flux using the semi-empirical Gr\"un relation \citep{Grun1985, Moorhead2020_MEM3_NTRS}, which underlies \texttt{MEM~3}’s baseline interplanetary meteoroid environment model. The Gr\"un relation describes the cumulative flux of meteoroids with mass greater than $m_{\rm min}$ (in grams) as
\begin{equation} \label{eq:grun}
    F_{\text{Gr\"un}}(>m_{\rm min}) = \big[A(m_{\rm min}) + B(m_{\rm min}) + C(m_{\rm min})\big]\, t_{\mathrm{yr}},
\end{equation}
where $t_{\mathrm{yr}} = 3.154\times10^{7}$~s is the number of seconds in one year, and
\begin{align}
    A(m_{\rm min}) &= (c_4\,m_{\rm min}^{\gamma_4} + c_5)^{\gamma_5}, \\
    B(m_{\rm min}) &= c_6\,\big(m_{\rm min} + c_7\,m_{\rm min}^{\gamma_6} + c_8\,m_{\rm min}^{\gamma_7}\big)^{\gamma_8}, \\
    C(m_{\rm min}) &= c_9\,\big(m_{\rm min} + c_{10}\,m_{\rm min}^{\gamma_9}\big)^{\gamma_{10}}.
\end{align}

The empirical coefficients and exponents are
\begin{align*}
    c_4 &= 2.2\times10^{3}, & \gamma_4 &= 0.306, \\
    c_5 &= 15,              & \gamma_5 &= -4.38, \\
    c_6 &= 1.3\times10^{-9},& \gamma_6 &= 2, \\
    c_7 &= 1\times10^{11},  & \gamma_7 &= 4, \\
    c_8 &= 1\times10^{27},  & \gamma_8 &= -0.36, \\
    c_9 &= 1.3\times10^{-16}, & \gamma_9 &= 2, \\
    c_{10} &= 1\times10^{6}, & \gamma_{10} &= -0.85.
\end{align*}

The Gr\"un relation is evaluated continuously over the range $10^{-6}\,\mathrm{g} < m_{\rm min} < 10^{1}\,\mathrm{g}$.  
Let $F_{\text{Gr\"un}}(>m_0)$ denote the cumulative Gr\"un flux at the anchor point $m_0 = 10^{-6}$~g, and $F(>m_0|\phi_{\rm south},\theta_{\rm south})$ the corresponding mean southern-pole impact rate from the \texttt{MEM~3} simulation at that mass threshold. We then define a scaled Gr\"un-based model for the impact rate near the south pole, $F(>m_{\rm min}|\phi_{\rm south},\theta_{\rm south})$, as
\begin{equation}
    \frac{F(>m_{\rm min}|\phi_{\rm south},\theta_{\rm south})}{ F(>m_0|\phi_{\rm south},\theta_{\rm south})} =
    \frac{F_{\text{Gr\"un}}(>m_{\rm min})}{F_{\text{Gr\"un}}(>m_0)}.
\end{equation}

This scaling preserves the functional shape of the Gr\"un mass--flux relation while normalizing it to the absolute impact rate derived from \texttt{MEM~3} at the south pole -- and allows us to predict $F(>m_{\rm min}|\phi_{\rm south},\theta_{\rm south})$ for any $m_{\rm min}$. The resulting smoothed relation, shown in Figure~\ref{fig: grun}, provides a continuous estimate of the expected meteoroid flux as a function of particle mass, which can be directly applied to evaluate shielding performance for various design thresholds. We find this scaled Gr\"un-based model agrees with the impact rates derived directly from our \texttt{MEM~3} runs (shown with black circles) as is expected.

This provides a useful tool for the community to evaluate whether a given shielding configuration meets a specified design threshold for acceptable penetrating impact rates, expressed as a function of shielding critical mass.

\section{Conclusion}
In this study, we used NASA’s Meteoroid Engineering Model~3 (\texttt{MEM~3}) to quantify the micrometeoroid impact environment across the lunar surface and evaluate its implications for long-duration Artemis-era surface habitats. By performing 1,000 directional \texttt{MEM~3} simulations uniformly distributed in selenographic coordinates, we derived both incident and penetrating impact rates for a notional lunar base with dimensions comparable to the International Space Station.

Our simulations indicate that a base of this size would experience approximately $15,000$ to $23,000$ incident micrometeoroid ($10^{-6}$--$10^{1}$~g) impacts per year, with the lunar poles receiving roughly 1.6 times fewer impacts than the equatorial regions. This longitudinal dependence is likely driven primarily by the geometric effects of the Moon’s orientation relative to the meteoroid sources and the partial gravitational focusing of fluxes by Earth.

Using a Whipple-type shield configuration and conservative assumptions for all parameters, we derived a critical projectile mass of $m_{\mathrm{crit}} \approx 10^{-1.16}$~g. Simulated impactor mass distributions show that approximately 99.9997\% of particles fall below this threshold, implying that current MMOD shielding technology can effectively mitigate nearly all micrometeoroid impacts. Accounting for modern Whipple shielding capabilities, we estimate a residual penetrating impact rate of $\sim$0.024--0.037 impacts per year, corresponding to a single penetrating impact event every 27--42~years. We again find that in this regime the lunar poles receive approximately 1.6 times fewer impacts than the equatorial regions.

To extend these discrete simulations, we scaled the empirical Gr\"un meteoroid flux relation to the \texttt{MEM~3}-derived impact rate at $m = 10^{-6}$~g, yielding a smooth function for the expected penetrating impact frequency as a function of limiting mass. This approach enables a continuous evaluation of shielding performance across arbitrary design thresholds and provides a practical engineering tool for mission planners.

Overall, our results demonstrate that:
\begin{enumerate}
    \item The lunar south pole offers a natural reduction in impact risk relative to equatorial sites, supporting its selection for sustained human presence.
    \item Gravitational focusing by Earth dominates over its geometric shielding of the micrometeoroid flux, making the net flux enhancement the primary Earth–Moon coupling effect relevant for surface impact risk in this mass regime.
    \item Existing Whipple shielding technology is sufficient to suppress micrometeoroid hazards by nearly five orders of magnitude, reducing the effective risk to a manageable level for current habitat designs.
\end{enumerate}

Future work should incorporate additional factors such as regolith-based or hybrid shielding materials, transient meteoroid streams, and local topographic effects on flux anisotropy. These refinements will further improve our understanding of the meteoroid threat environment for upcoming Artemis surface missions and long-term lunar infrastructure.

\section*{Acknowledgments}

We are very grateful to William Bottke, David Kipping, and David Nesvorn\'y for comments and feedback on the manuscript. We would also like the thank Ben Cassese and members of the Astronomical Data Group at the Flatiron Institute, Center for Computational Astrophysics, for fruitful discussions. This project started as a part of Michael Massimino's \textit{Introduction to Human Spaceflight} class in the spring of 2022 at Columbia, and we are very thankful for his help and guidance during its initial development.

N.A., M.D., and K.O. were supported by the Student Training in Astronomy Research (STAR) program at Columbia University, which is grateful for the support of the Pinkerton Foundation, New York City Science Research Mentoring Consortium, and the National Osterbrock Leadership Program of the AAS.

D.A.Y. acknowledges support from NASA Grant \#80NSSC21K0960. D.A.Y. acknowledges support from the NASA/NY Space Grant. D.A.Y. thanks the LSST-DA Data Science Fellowship Program, which is funded by LSST-DA, the Brinson Foundation, the WoodNext Foundation, and the Research Corporation for Science Advancement Foundation; his participation in the program has benefited this work.

Special thanks to donors to the Cool Worlds Lab, without whom this kind of research would not be possible: Douglas Daughaday,
Elena West,
Tristan Zajonc,
Alex de Vaal,
Mark Elliott,
Stephen Lee,
Zachary Danielson,
Chad Souter,
Marcus Gillette,
Jason Rockett,
Tom Donkin,
Andrew Schoen,
Mike Hedlund,
Ryan Provost,
Nicholas De Haan,
Emerson Garland,
Queen Rd Fnd Inc.,
Ieuan Williams,
Axel Nimmerjahn,
Brian Cartmell,
Guillaume Le Saint,
Robin Raszka,
Bas van Gaalen,
Josh Alley,
Drew Aron,
Warren Smith,
Brad Bueche,
Steve Larter,
Marisol Adler \&
Craig Frederick.

The Meteoroid Engineering Model is publicly available from NASA's Meteoroid Environment Office upon request (\url{https://www.nasa.gov/offices/meo/software/mem_detail.html}).

This work made use of the following software packages: \texttt{astropy} \citep{astropy:2013, astropy:2018, astropy:2022}, \texttt{matplotlib} \citep{Hunter:2007}, \texttt{MEM~3} \citep{Moorhead2020_MEM3_NTRS}, \texttt{numpy} \citep{numpy}, \texttt{python} \citep{python}, and \texttt{scipy} \citep{2020SciPy-NMeth, scipy_17101542}.

Software citation information aggregated using \texttt{\href{https://www.tomwagg.com/software-citation-station/}{The Software Citation Station}} \citep{software-citation-station-paper, software-citation-station-zenodo}.

\texttt{ChatGPT} was
utilized to improve wording at the sentence level and assist with
coding inquires -- last accessed in 2025 November.

%





\newpage
\bibliography{main}{}
\bibliographystyle{aasjournal}




\end{document}